\documentclass[epjST]{svjour}
\usepackage{graphicx}
\usepackage[numbers,compress]{natbib}
\usepackage{amsmath}
\usepackage{amssymb}
\usepackage{hyperref}
\usepackage{graphics}

\begin{document}
\title{Averaged Recurrence Quantification Analysis}
\subtitle{Method omitting the recurrence threshold choice}
\author{Radim Pánis\inst{1,3}\fnmsep\thanks{\email{radim.panis@physics.slu.cz}} \and Karel Adámek \inst{1,2} \and Norbert Marwan \inst{3}}
\institute{Research Centre for Theoretical Physics and Astrophysics, Institute of Physics, Silesian University in Opava, Bezru{\v c}ovo n{\'a}m.13, CZ-74601 Opava, Czech Republic
\and Oxford e-Research Centre, Department of Engineering Science, University of Oxford, 7 Keble Rd,
Oxford OX1 3QG, United Kingdom \and Potsdam Institute for Climate Impact Research (PIK), Member of the Leibniz Association,
Potsdam 14412, Germany}
\abstract{
Recurrence quantification analysis (RQA) is a well established method of nonlinear data analysis. In this work we present a new strategy for an almost parameter-free RQA. 
The approach finally omits the choice of the threshold parameter by calculating the RQA measures for a range of thresholds (in fact recurrence rates). Specifically, we test the ability of the RQA   measure determinism, to sort data with respect to their signal to noise ratios. We consider  a periodic signal, simple chaotic  logistic equation, and Lorenz system in the tested data set with different and even very small signal to noise ratios of lengths  $10^2, 10^3, 10^4,$ and $10^5$. To make the calculations possible a new effective algorithm was developed for streamlining of the numerical operations on Graphics Processing Unit (GPU).
} %end of abstract
\maketitle

\section{Introduction}

The authors in \cite{2015Chaos..25i7610B} wrote in the very end: ``Will it be possible to design algorithms
whose free parameters can be chosen systematically, via
intuition, or perhaps even automatically? Such developments would streamline nonlinear time-series analysis,
making it an indispensible tool to make sense out of the
real world.''
In this article we present a way of  reducing the effect of a specific choice of the important threshold parameter in the well established nonlinear method of recurrence quantification analysis (RQA).

The popularity of RQA is continuously increasing in many fields across the science spectra -- physical, biological, economical, simply everywhere where data can be obtained\cite{2015rqa..book.....W}.

There is yet an open question on the selection of the basic parameter needed for the calculation of the recurrence plot (RP), the base of the RQA. The RP needs for its calculations a threshold parameter, $\varepsilon$, those selection can influence the results but depends on the 
specific research question\cite{marwan2011}.
The choice of  $\varepsilon$ was already discussed in previous studies \cite{2002PhyD..171..138T,2021Chaos..31h3131M,2020Chaos..30a3124A,2008EPJST.164...45S}.

In this work we propose and study the possibility of providing RQA in a different way by omitting the choice of  single $\varepsilon$ and, thus, making RQA more stable and objective.
Here we use RQA for comparing data with different amounts of noise, where the deterministic behaviour in time series can be reflected at more scales \cite{2000PhRvE..62..427C}, corresponding, e.g., to different thresholds $\varepsilon$. 
Therefore, we use a set of $\varepsilon$ values corresponding to a given set of recurrence rates (RRs) while we test the assumption that averaging the RQA resulting by the selected RRs improves the analysis.
We provide the testing for various types of data  of different lengths  to study also the relation with the amount of data points needed in order to sort them according to its deterministic content.
The data have different signal-to-noise ratios while we focus on the RQA measure determinism (DET) in order to explore the ability to recognize deterministic structures and evaluate quantitatively
the amount of contaminating noise.

In order for the numerical tests to be carried out in an acceptable amount of time, we perform the RQA calculation on a Graphics Processing Unit (GPU) which also significantly enlarges the possible dimensions of the input time series for RQA due to the effective memory handling.
  This new computation method opens the door for  desktop computers to perform RQA for large data  alongside with the proposed novel  approach to deliver more robust results.

When analysing time series from experiments or real world applications, it is difficult
to distinguish between deterministic chaos and stochastic behaviour due to the finite
length of the data and the different intrinsic scales\cite{2000PhRvE..62..427C}. 
An example of such an application those variability is analysed at more scales is
data originating from extragalactic sources. Their data is  varying from minutes to several decades and  details of the processes leading to such  multi-timescale variability are still under discussion \cite{1996ASPC..110..391U, Mohorian_2021, Mayer2010}. In addition the data originating in extragalactic sources are affected by noise \cite{1991ApJ...373..465F}, where the signal-to-noise ratios (SNR) tend to be very small (an extreme example are gravitational waves \cite{1995pnac.conf..160T}).

Recurrence quantification analysis (RQA) is a promising tool to distinguish different types of dynamics, such as from deterministic chaos and noise \cite{2002PhRvE..66b6702M,2008EPJST.164...45S}. 
In contrast to the usual approach, we do not use one fixed value of recurrence threshold $\varepsilon$ for the RQA calculation, but we 
evaluate the behaviour of the RQA measures for a range of thresholds.
The assumption is that this approach will reveal the true dynamics of the underlying system  at more scales. Thus, the main goals of this study is testing this approach
by evaluating the results and comparing them with the results by a single choice of $\varepsilon$.

Taking into account the variability of the measured data of some unknown physical system, the more in context of deterministic chaos, nonlinear phenomena or even complex systems  which produce time series, the amount of uncertainty is high and the data often resemble random numbers. We can get the feeling, the choice of input parameters for the  algorithms  gives quite biased result.  This fact often causes struggles in any numerical data analysis.
Techniques and algorithms which reduce the number of parameters bring in some way releasing feeling of the unbiased result by the person using it, a good example is, e.g., finding embedding parameters \cite{kraemer2021}.
Another intuitive  reason can be identified within this context -- the less free parameters the algorithms, equations, or models require, the closer we move to the ground physical theories, laws of nature, expressed by elegant formulas.

%where the good example of such doing can be found by the creation of the Einsteins gravitational law.  Einstein did not use any numerical solutions for its creation, so as there is some contempt about numerics from the view of the pure theoreticians, but he did identify the right.

In Sects.~\ref{s1} and \ref{s2} we provide brief explanation of RQA with the description of the averaging approach.  In Sect.~\ref{s3} we describe the testing of the new approach and in
 \ref{s4} the technical implementation of the algorithm for GPUs. Finally, the results are presented in Sect.~\ref{s5} and discussed in \ref{s6}.

\begin{figure}[!htb] 
\begin{center}
\includegraphics[width=.9\linewidth]{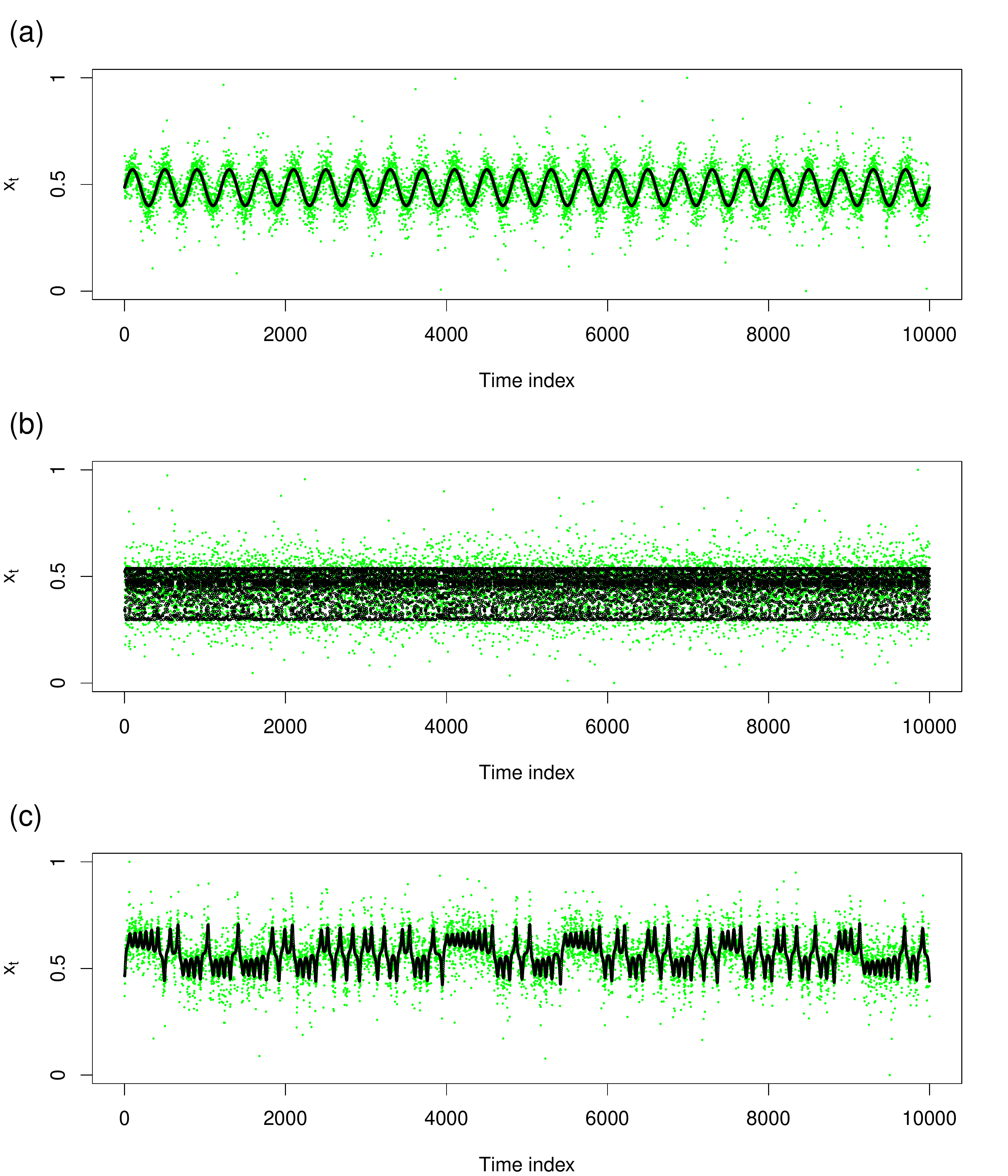}
\caption{The examples of studied time series, (a) periodic function, (b) Logistic map, (c) Lorenz system, the green points denote the time series disturbed with white noise, the black points denote the time series without noise. Here the SNR  is 1:1.}
\label{f1}
\end{center}
\end{figure}

\section{Recurrence quantification analysis \label{s1}}
Recurrence quantification analysis (RQA) is  a handy and versatile tool of nonlinear analysis, introduced in 1992 by \citet{1992PhLA..171..199Z} and extended by \citet{2002PhRvE..66b6702M}.
RQA provides measures of complexity that evaluate the properties of the recurrence plot (RP), a graphical tool used for investigating the behaviour of state space trajectories $\vec{x}_i$\cite{1987EL......4..973E}. 
  
The basis of RQA is calculating the recurrence matrix 
\begin{equation}  
R_{i,j}= H(\varepsilon - \| \vec{x}_i - \vec{x}_j\|  ) \quad  i, j = 1, ...,N,   \label{Eq:1}
\end{equation}
where $N$ is the number of measured points $\vec{x}_i$, $ \| \cdot \|$ is a norm which, in this work, is  the maximum norm 
$\| \vec{x} \|_{\max}: = \max(| x_1|, \dots ,|x_n |)$ for $\vec{x}\in \mathbb{R} ^{n}$.
$\varepsilon$ is a threshold distance, defining the recurrence of a state by its spatial closeness to a former state. The selection of $\varepsilon$ is crucial and depends on the specific research
question and can have a strong effect on the result. $H(\cdot)$, is the Heaviside function,  defined as 
\begin{equation}
H(\varepsilon) =
    \begin{cases}
            1, &         \varepsilon > 0\\
            0, &     \varepsilon \leq  0.
    \end{cases}
    \label{Eq:2}
\end{equation}
  
 Eq.~(\ref{Eq:1}) results in a symmetrical square matrix that consists of binary values, i.e., zeros and ones. The RP is obtained as a plot of this square matrix. 
 
 There RQA measures we use in this work are
\begin{enumerate}
 \item Recurrence rate (RR) 
\begin{equation}
RR = \frac{100}{N^2} \sum_{i,j=1}^N R_{i,j}  \label{Eq:3},
\end{equation}
 which provides a measure for the density of recurrence points in the RP. Henceforth, we express this measure in percentages instead of decimals. $RR$ is directly related
 to the threshold $\varepsilon$ and is often used as the alternative way to
 preselect $\varepsilon$\cite{kraemer2018}.

 \item Determinism (DET), quantifying how deterministic or well behaved a system is
\begin{equation}
DET  = \frac{ \sum_{l=l_\text{min}}^N l P(l) }{ \sum_{i,j=1}^N R_{i,j} }, \label{Eq:4}
\end{equation}
where  $P(l)$ denotes the frequency distribution of the lengths $l$ of the diagonal lines and $l_{\min}$ denotes the minimal amount of points considered as line and it is set up  to 2 as minimal value for all the calculations within this study.
\end{enumerate}

 Diagonal lines of the RP,
parallel to the main diagonal point to the joint period when a trajectory accompanies locally close paths. Therefore, diagonal lines in the RP present the information of predictability and the deterministic content  of the dynamical system. This property naturally suggest that the DET measure should be able to distinguish between signal of deterministic origin and stochastic noise.

\section{Averaged RQA \label{s2}}

As mentioned above, we consider a range of thresholds $\varepsilon$ to cover all scales in the
variability of the time series. Instead of setting $\varepsilon$ to a certain value,
we preselect $RR$ and use the corresponding value for $\varepsilon$\cite{kraemer2018}.

The novel approach here is to use averaged RQA quantities, which are calculated 
for a range of $RR$ values. For example, 
for the measure of determinism we define the {\it averaged determinism} 
\begin{equation}\label{eq_avgDET}
\overline{DET}_{RR_*} = \frac{1}{n} \sum_{RR=1}^{RR_*} DET_{RR},
\end{equation}
where the  $DET_{RR}$ denotes the $DET$ corresponding to a given value of the measure $RR$, $n$ is the number of considered $RR$ values for averaging, and $RR_*$ means the highest $RR$ to which is averaged and it can take any value from the interval $[0,\ldots, 100]$. Naturally, the case $n = 1 $ cannot be seen as averaging and the case $RR_* = 100 $ should not be included, because the resulting RP does not contain any useful information about the dynamics.
In this work we test the averaging for $ RR_* \in [1,2,3, \dots, 99]$.

% The process of searching for the $\varepsilon$-values  corresponding to the set of $RR \in [1,2,3 \dots, RR_*][\%]$ for a given time series  $x_t$ is one of the important points of this work.  
% Naturally, the dynamics of arbitrary time series does not guarantee that, the $\varepsilon$ values   corresponding to the set of $RR \in [1,2,3 \dots, 99][\%]$ can be found in a way of achieving this integer like precision of the $RR$ values. The error can be simply defined as the difference between the $RR$ corresponding to some  the $\varepsilon$ which we were searching for and its desired integer value, when expressed in \%. 
%We, for the purpose of achieving  small errors of the $RR$ values have implemented the approach of calculating of the $RR$ for 400 thresholds for every time series and then selecting the 99 closest  ones  to the desired values of $RR \in [1,2,3 \dots, 99][\%]$. In order to use one fixed set of thresholds, every time series was transformed into interval [0,1].
%In this work, we provided this calculation for 3600 time series, see \ref{s3}, and the error for any $RR$ (from $3600*99 = 356400$ ) was on average $ \approx  0.5 \%$, while the maximal error was $ \approx  1 \%$.
% In order to achieve this precision, the implementation of the algorithms on GPU \ref{s4} was essential, while the whole computation took roughly 5 days.

Naturally, this approach can also be applied for other RQA measures\cite{2015rqa..book.....W} in order to catch their behaviour on more scales. This idea somehow resembles the concept of unthresholded recurrence plots discussed, e.g., by \citet{1998Chaos...8..861I}, where instead of constructing a RP by thresholding the pairwise distances $\| \vec{x}_i - \vec{x}_j\|$ and representing the matrix by two colours, the entire distance matrix is represented, encoding the nonlinear properties of the system on all scales.

\section{Methodology \label{s3}}

The application of the above approach is used for discriminating different levels of
noise from a deterministic signal. To illustrate this, we generate data with different fixed lengths, namely $10^2, 10^3, 10^4,$ and $10^5$, where the number is in the sense of density of points, iteration, or integration time (divided by 100, because the step of 0.01) for periodic function, logistic map, and Lorenz system, respectively (details in Subsect.~\ref{ss1}, \ref{ss2}, and \ref{ss3}). In order to investigate the quantitative information for every representative we generate 10 data sets of equidistantly different SNRs,  where  SNR is defined as the ratio between signal variance and noise variance.

We are studying the ability of sorting according to the various SNRs (definition below), expressing the ability to sort the 10 signals with different SNRs, while it is averaged for 10 realizations. 
In addition, we generate  the  equidistant SNR ratios in 3 intervals to inspect the boundaries of this approach, namely, $[0.01,0.02,\dots ,0.1]$, $[0.1,0.2,\dots ,1]$, and $[1,2,\dots ,10]$, where for each SNR interval there are 10 equidistantly separated values   denoted as ``SNR L.~0.1'',  ``SNR L.~1'', and ``SNR L.~10'' respectively.

 The lowest SNR  in this work analysed is of the order of one hundred  of signal to one, or another way expressed 1:100 (signal:noise) and the highest SNR in this set is of the ratio 10:1.
Overall the testing has been done on 3,600 time series, what corresponds to: 3 types of dynamics (periodic, Logistic map, Lorenz system) $\times$ 3 SNR Levels (SNR L.~= 0.1, 1, and 10) $\times$ 4 lengths ($10^2, 10^3, 10^4, 10^5$) $\times$ 10 generated time series corresponding to some SNR Level  $\times$ 10 realizations.

As a measure evaluate the sorting of time series according to their deterministic content (estimated by $\overline{DET}_{RR_*}$ or $DET_{RR}$) we introduce the sorting rate $S$.
The sorting rate is defined as the difference between the places in ordering of $n=10$ time series by a $\overline{DET}_{RR_*}$ or $DET_{RR}$ measure,  expressed by  vector $x$  and the vector of the defined positions $y$ in absolute value, summed and then expressed as percentage of successful sort. It can mathematically be
expressed as
\begin{equation}
S = \frac{\sum_{i=1}^{10} \left(\sum_{j=1}^{10} |z_j - y_j|\right)_i - \sum_{i=1}^{10} \left(\sum_{j=1}^{10} |x_j - y_j|\right)_i}
{\sum_{i=1}^{10} \left(\sum_{j=1}^{10} |z_j - y_j|\right)_i} \label{Eq:5}
\end{equation}
where $x_j$ is the $j$-th element of the vector $x$ for some $i$-th realization, denoting the SNR/position of time series according it’s $\overline{DET}_{RR_*}$ or $DET_{RR}$ measure, and $y_j$
is the element of the vector of $y$ of the SNRs/positions in the natural (sorted) way, i.e., $y =  [1, 2, 3, 4, 5, 6, 7, 8, 9, 10] $, and vector  $z$ represents the reversed $y$ vector.
The summation over index $i$  is in the sense of the repetition of the same experiment, or in other words for 10 realizations of corrupting the time series by white noise, for the sake of obtaining a stable result.
The term $\sum_{i=1}^{10} (\sum_{j=1}^{10} |x_j - y_j|)_i)$ can be seen as the departure from the best sorting, and $\sum_{i=1}^{10} (\sum_{j=1}^{10} |z_j - y_j|)_i)$ takes naturally the value of  500 when $i,j = 1,2,\dots, 10$ and  represents the worst possible scenario. In the following, $S$ is expressed as percentages instead of decimals. A value of 100\% means
a successful sorting, a value towards 0\% means a complete failure of sorting.

%One can observe how this measures works in Table B3, which shows that in this particular case, the. Consequently, we consider the10 particular SNRs, the measures for which are bounded in the intervals, where measure1 

\subsection{Periodic time series \label{ss1} }
In order to simulate periodic signals we have used the R library RobPer \cite{robper} which has the function tsgen, originally made for simulating light curves.
The function actually has 11 parameters and allows to simulate periodic signals, which mimic the data from observations thanks to the features, e.g., presence of outliers or gaps. 
We set up parameter of sampling  to  ``equi'' for equidistant sampling without gaps and the  type of the periodic fluctuation to ``sine''. The  number of sampling cycles that is observed is set up to 25 (Fig.~\ref{f1}a).

\subsection{Logistic map \label{ss2}} 
The Logistic map is a classical example of a simple non-linear dynamical system  exhibiting a variety of periodic and chaotic dynamics, given by the quadratic equation
$x_{n+1} = r x_n (1 -  x_n)$.
For the initial value of the variable $x_0 \in (0, 1)$, the logistic map  generates sequences of real numbers $x_n \in (0, 1)$. The behaviour of the sequence $x_n$ depends on the parameter $r$.   Roughly speaking, the behaviour on $ r \in [r_0, 4]$  is chaotic, with some occasional ``islands of regularity''. The transition between regular and chaotic behaviour happens for  $r = r_0 \approx 3.56995$. 
In this work we generate time series from the logistic map  using $r = 3.679 $, where the band merging causes frequent laminar states \cite{2002PhRvE..66b6702M} (Fig.~\ref{f1}b).

\subsection{Lorenz system \label{ss3}} 
The well-known Lorenz system  is continuous nonlinear, non-periodic, three-dimensional, and deterministic. 
The famous  attractor can  can be reproduced by
solving $\dot{x} = \sigma (y - x)$, $\dot{y} =  x (\rho - z)$, and  $\dot{z} =  xy - \beta z)$,  with $\sigma = 10$, $\beta = 8/3$, and $\rho = 28$. The equations are integrated numerically with  a Runge-Kutta solver and a time step 0.01. Finally, we use the $x$ value to emulate an observation by just one time series  (Fig.~\ref{f1}c).

An essential step in nonlinear time series analysis is state space reconstruction.
The dynamics of a $m$-dimensional nonlinear system can be reconstructed (in topological sense) from a single time series using the mathematical embedding theorem \cite{1981LNM...898..366T}.
The usual approach of state space reconstruction is  delay coordinate embedding. The original scalar  time series is mapped into a new space which is defined by the number of delayed dimensions $m$.
The $m$-dimensional  vector  $\vec{x}(t)$ is constructed from $m$ samples of time series ${y}(t)$ using the delay $\tau$ by $
 \vec{x}(t) =  [{y}(t), {y}(t-\tau), {y}(t-2\tau), \dots, {y}(t-(m-1)\tau)  ]$
In practice, when dealing with unknown systems, the values of $\tau$  and $m$ need to be estimated numerically\cite{2015Chaos..25i7610B,1997PhyD..110...43C, 2017arXiv170405199J}. However, in the case of  Lorenz system the dynamics is known and the parameters are set up to $\tau = 3$  and $m = 3$.

\section{Technical implementation on a GPU} \label{s4}

The measures of RQA are computationally
 expensive when computed naively because they are
 calculated from a RP, Eq.~(\ref{Eq:1})
 that grows as $\mathcal{O}(N^2)$, where $N$ is the length of the time series (or phase space vector) being analysed. More importantly, the memory footprint also grows as $\mathcal{O}(N^2)$. Thus, even modestly sized time series will take a long time to be calculated on a standard system. Therefore, developing faster implementations and techniques to calculate RQA measures is crucial. \citet{SCHULTZ2015997} have shown that in the special case where the threshold is zero, some RQA measures can be obtained with $\mathcal{O}(N \log(N))$
complexity in $\mathcal{O}(N)$ space and have proposed approximations for RQA measures that have same computational complexity for thresholds above zero \cite{SCHULTZ2015997}.

Calculating the RQA measures using a parallel approach where we can distribute the computational load to multiple processors/cores is equally important. GPUs are an ideal platform for RQA implementation as they possess a good combination of a large number of processing cores (NVIDIA A100 GPU has 6912 floating-point cores) and high bandwidth to memory (NVIDIA A100 has 1555GB/s). We have developed a parallel GPU accelerated software for NVIDIA GPUs written in CUDA. There are other packages which take advantage of parallelising calculations and GPU acceleration. For example, \citet{Rawald2017PyRQA} have implemented RQA calculation using the OpenCL framework. We have designed the algorithms to have a minimal memory footprint ($\mathcal{O}(N)$) to allow performing RQA even on very long time series (100k+ points).

For most RQA measures, two types of tasks are required. First, counting the frequency of lines $l$ of a given length, producing a histogram of diagonal line lengths. The second is the density of the RP, which is used to calculate RR. However, RR can also be calculated from the histogram of line lengths. Thus, the histogram is used for the calculation of (DET, L, $\mathrm{L}_{\max}$ and RR). To calculate the histogram, we exploit the symmetric property of the $R_{i,j}$ matrix that allows us to use only the upper triangle of the $R_{i,j}$ matrix (i.e., $j>i$). We also calculate the value of the element $R_{i,j}$ on the fly, thus avoiding a significant memory footprint that would be otherwise required to store the whole $R_{i,j}$ matrix or any of its sub-matrices.

To calculate the histogram of line lengths, we use a stencil operation (a filter applied at every point of the RP) that flags the beginning and the end of each line. These flags are then aligned and compared, allowing us to calculate the length of all lines in a parallel implementation on multiple GPU workers. 
%These are then added to the histogram of lengths. These stencil operations allow us to evaluate the position of all line beginnings and line ends independently and in a parallel fashion. 

\section{Results \label{s5}}

We focus on a comparison between the averaging approach and the approach of a fixed choice of some threshold value.  
%for this purpose we present the table \ref{tab:table1} where the comparison between 2 specific  pairs of the RR values is present, the figures \ref{f2}, \ref{f3} and \ref{f4}, where the results for all possible choices of RR are depicted graphically, the figure \ref{f5} of averaged quantities  and the table \ref{tab:table2}, which were made in purpose to reveal some potential pattern of preferred choice of RR.

 In Tab.~\ref{tab:table1} the sorting rates $S$ between the $\overline{DET}_{99}$  and the $DET_1$ are presented, what corresponds to the average of
 $DET$ for  $ RR \in [1,2,3, \dots, 99][\%]$
and just for $RR =  1[\%]$, respectively. Thus, $\overline{DET}_{99}$ is the universal choice which covers all the scales (except for RR=100\%). $DET_1$ is selected for this comparison as the value of RR as recommended by \citet{2002PhLA..297..173Z} for the construction of RPs.
Here we find that the choice of $DET_1$
performs better only in few cases when the lengths and SNRs of the time series are lower (Tab.~\ref{tab:table1}).

\begin{figure*}[!htb] 
		\centering 
	\includegraphics[width=1\linewidth]{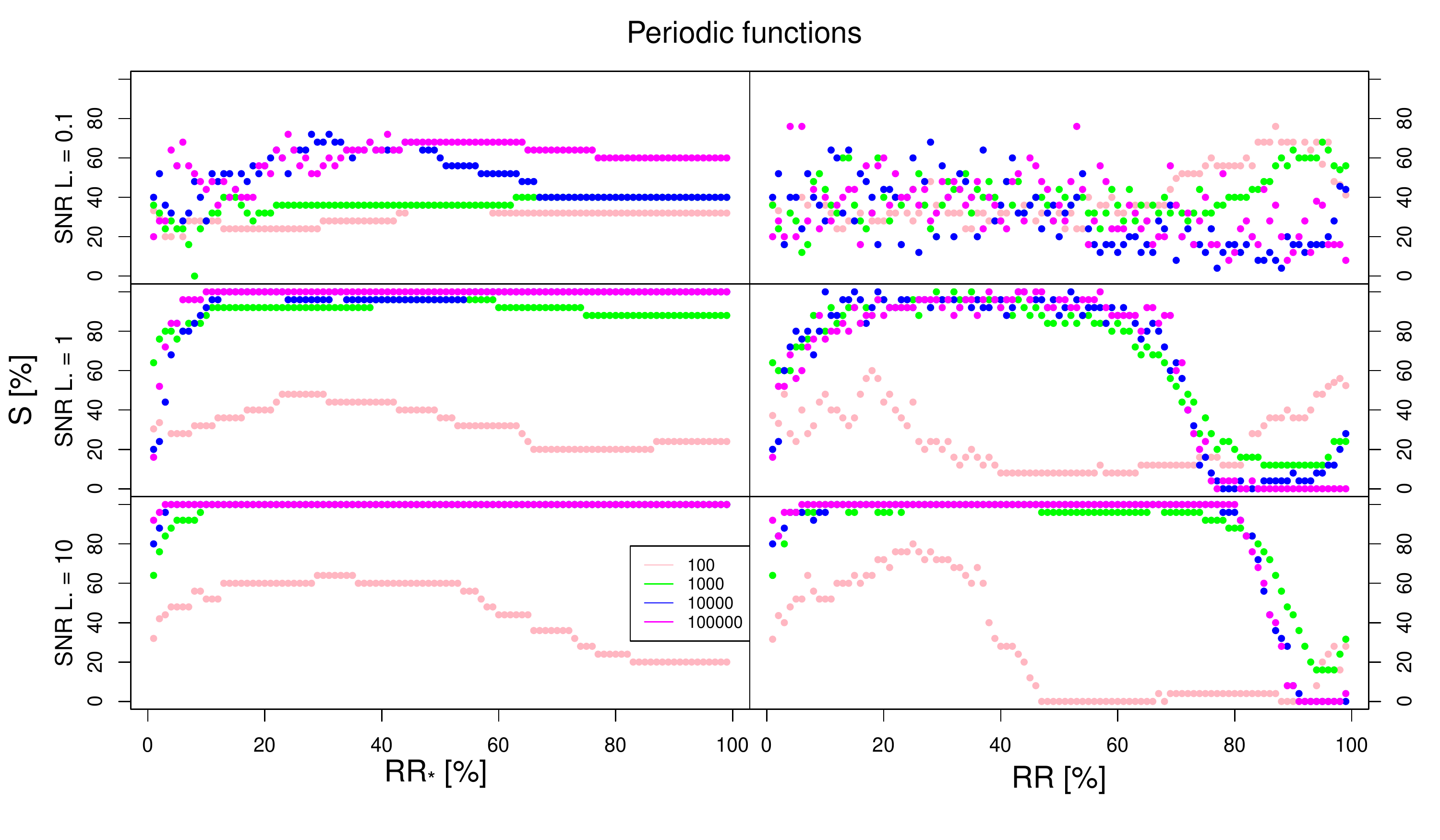}
\caption{\label{f2} This plot shows the sorting rate $S$ of both averaging and classical approach of choosing $RR_*$ or $RR$ for the disturbed periodic functions. }			
\end{figure*}

\begin{figure*}[!htb] 
		\centering 
	\includegraphics[width=1\linewidth]{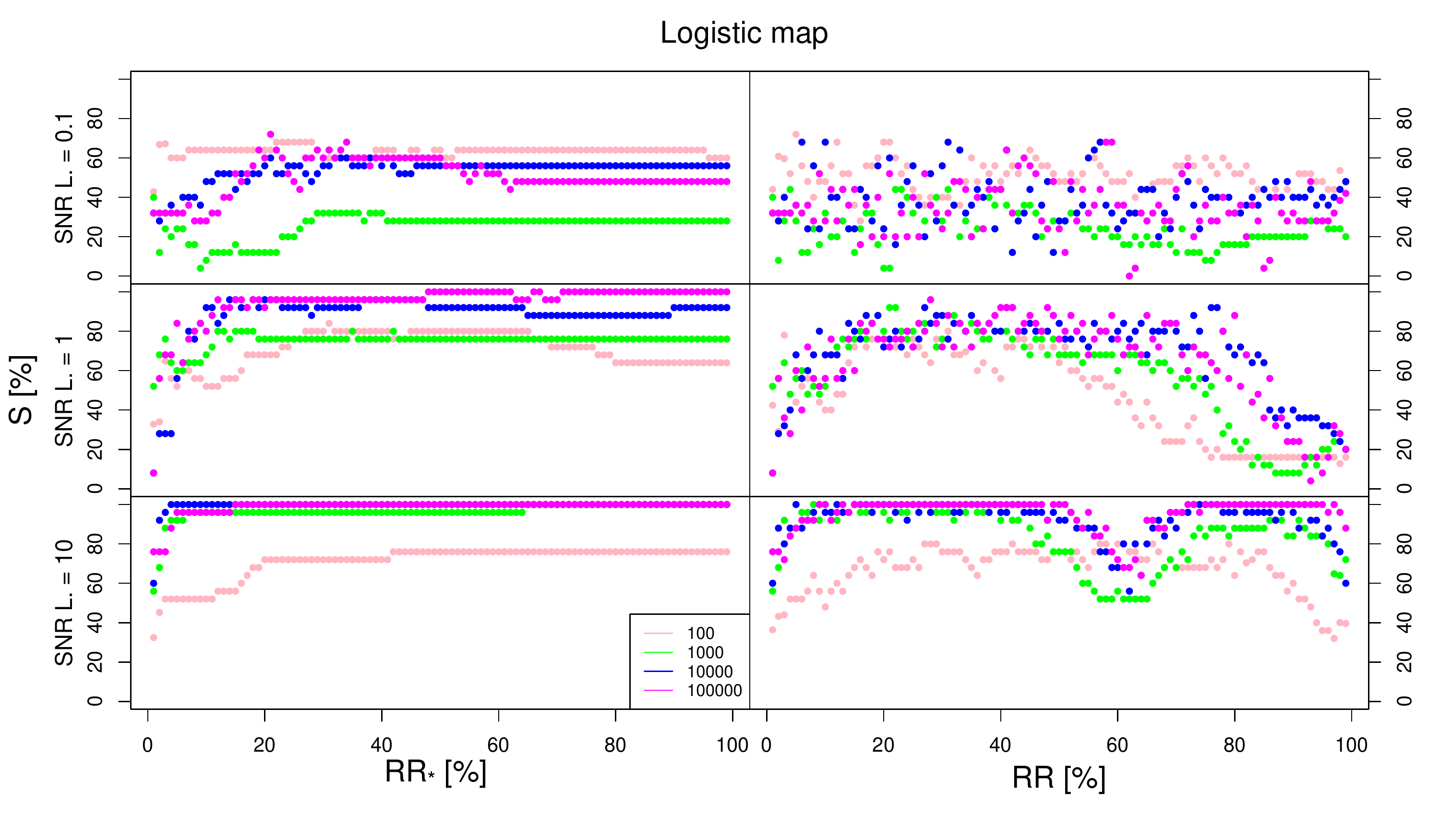}
\caption{\label{f3} This plot shows the sorting rate $S$ of both averaging and classical approach of choosing $RR_*$ or $RR$  for the disturbed Logistic map. }			
\end{figure*}

\begin{figure*}[!htb] 
		\centering 
		\includegraphics[width=1\linewidth]{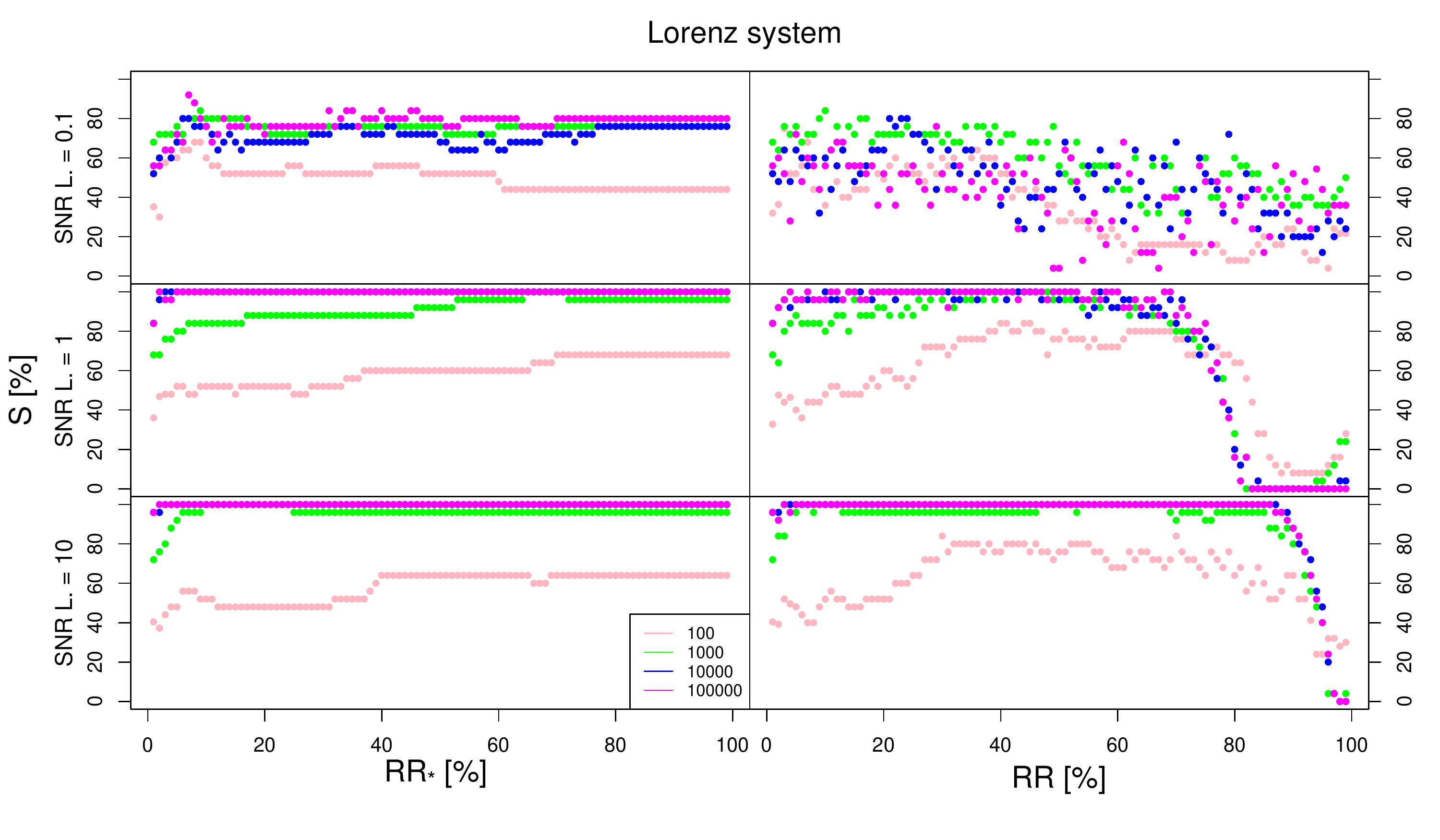}
\caption{\label{f4} This plot shows the sorting rate $S$ of both averaging and classical approach of choosing $RR_*$ or $RR$  for the disturbed Lorenz system. }			
\end{figure*}

Next we consider the performance of the sorting using $\overline{DET}_{RR_*}$ for different values of $RR_*$ as the largest limit in Eq.~(\ref{eq_avgDET}) and for the single RR based $DET_{RR}$ (Figs.~\ref{f2}, \ref{f3} and \ref{f4}).
We find more robust results by the averaging approach $\overline{DET}_{RR_*}$ when compared to $DET_{RR}$. 

For better understanding, the results from the first row in Tab.~\ref{tab:table1} are depicted in Fig.~\ref{f2}  as the first left-right pair from the top where the $y$-axis is denoted as ``SNR L.~$ = 0.1$''. The colours correspond to the considered time series length. 
Following this logic, the values 32 / 33.2  (Tab.~\ref{tab:table1}) representing $S$ for the periodic functions with the  level of $SNR \in [0.01,0.02,\dots ,0.1]$ of the lengths 100  depicted in Fig.~\ref{f2},  are 
$\overline{DET}_{99}$, which is visible as the last point of the light brown line in the left plot and the $DET_1$ value,  which naturally is to observe in both parts of the figure  because of $\overline{DET}_{1} = DET_1$, as the average of one value is the value itself.

 %The comparison of the all 3 graphics says that that, for all 3 tested generators in the averaging approach, namely sine functions, logistic map, and Lorenz system, the ability to sort the data according to the SNR grows with the amount of data analysed, as the purple curve is almost all the time above the others, the order 

The most interesting feature of the comparison of the both approaches is the consistency of the averaging approach, for which the variability in $\overline{DET}_{RR_*}$ tends to be rather steady with low variance
after some values of $RR$ (Figs.~\ref{f2}, \ref{f3} and \ref{f4}, left panel), while for ${DET}_{RR}$ the ability to sort the data is in most of the cases not steady  for neighbouring $RRs$ (Figs.~\ref{f2}, \ref{f3} and \ref{f4}, right panel).

\begin{figure*}[htb] 
		\centering 
		\includegraphics[width=1\linewidth]{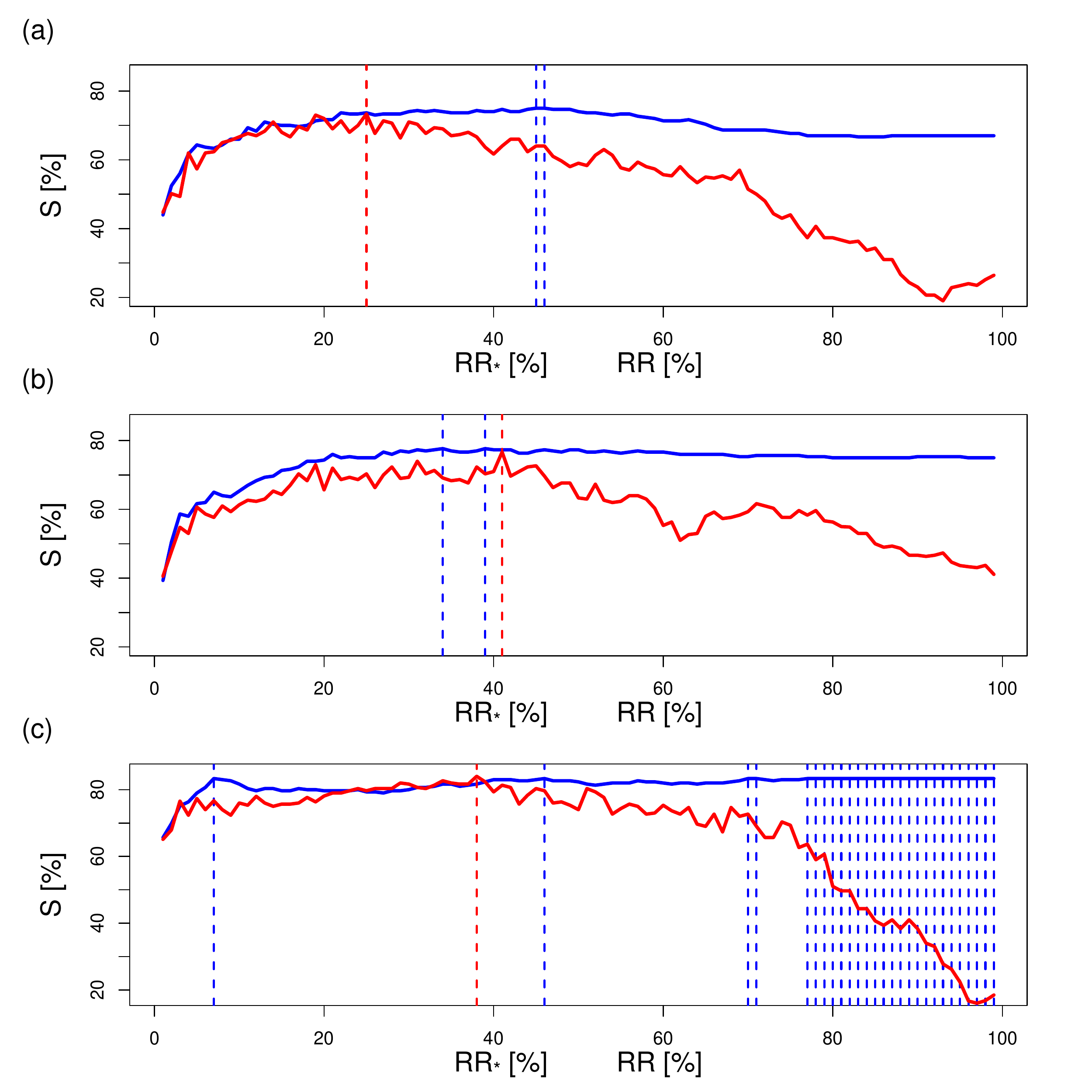}
\caption{\label{f5} Averaged sorting rate $S$ of the averaging approach (blue line) and fixed threshold approach (red line).  The dashed vertical lines denote the position of the maximum. The presented $S$ are averages over all lengths and SNR levels for (a) periodic functions, (b) logistic map, and (c) Lorenz system respectively.  The labels on y-axis denotes both approaches, averaging and fixed one respectively. We observe that for most of the $RR_*$ and $RR$ values, the blue line is above the red line, representing better sorting performance of the average determinism.}	
\end{figure*}

The simulations of the ability to sort data according their SNR  measured with the sorting rate $S$ is  in favour of the averaging approach, primarily in such  sense, when we chose some RR value, the averaging up to that RR results in a more accurate and robust result.
Although some exceptions can be found, the importance of the averaging is the consistency which is absent in the standard approach of choosing one fixed $RR$. This numerical simulation also shows that there is no universal, preferred threshold (or RR) value for which the best results can be achieved (Tab.~\ref{tab:table2}, Fig.~\ref{f5}).
Here we further average $\overline{DET}_{RR_*}$ over all considered lengths and SNRs 
(as shown in Figs.~\ref{f2}, \ref{f3}, and \ref{f4}) to provide an aggregated
impression of the dependency of the accuracy with respect to the chosen maximal RR.
We observe some trends, e.g., both approaches perform better for $RR > 15\%$, which
was not so obvious in Figs.~\ref{f2}, \ref{f3}, and \ref{f4}.
On the other side of the sorting ability, the approach with single choice of RR is getting low for very high roughly $RR > 60\%$ where the thresholds are too high to recognize the determinism, while this phenomenon is sometimes also present for the averaging approach but to much less extent.

We observe that the ability to sort the time series according their SNR is mostly ordered according the length of analysed time series (Fig.~\ref{f2} to \ref{f4}),  the gap between the shortest time series and the rest is mostly visible by  the averaging approach, while for the single choice approach there is often no such clear pattern to observe.

\begin{table} 
\caption{\label{tab:table1} Sorting rate $S$ for various time series, SNRs and time series lengths. The number left of the slash denotes $S$ of the averaging approach, when averaged to $RR_* = 99\%$, the right sight the case for fixed $RR = 1\%$.}
\centering
\begin{tabular}{lllll}
  \hline
  type/length & 100 & 1000 & 10000 & 100000 \\ 
  \hline
  sin 0.1 & 32 / 33.2 & 40 / 36 & 40 / 40 & 60 / 20 \\ 
    sin 1 & 24 / 30.4 & 88 / 64 & 100 / 20 & 100 / 16 \\ 
   sin10 & 20 / 32 & 100 / 64 & 100 / 80 & 100 / 92 \\ 
   log 0.1 & 60 / 42.8 & 28 / 40 & 56 / 32 & 48 / 32 \\ 
   log 1 & 64 / 32.8 & 76 / 52 & 92 / 8 & 100 / 8 \\ 
   log10 & 76 / 32.4 & 100 / 56 & 100 / 60 & 100 / 76 \\ 
   lor 0.1 & 44 / 35.2 & 76 / 68 & 76 / 52 & 80 / 56 \\ 
   lor 1 & 68 / 36 & 96 / 68 & 100 / 84 & 100 / 84 \\ 
   lor 10 & 64 / 40.4 & 96 / 72 & 100 / 96 & 100 / 96 \\ 
   \hline
\end{tabular}
\end{table}

\begin{table*}[htbp]
\caption{\label{tab:table2}This table shows the $RR_*$ or $RR$, for which the best sorting is achieved.
The number to left of the slash denotes the $RR_*$  of the averaging approach for which the best sorting is achieved, the right sight the case for fixed $RR$, the numbers in parenthesis denote the amount of higher $RR_*$ or $RR$ for which the same sorting rate $S$ would be achieved.}
%\begin{ruledtabular}
\begin{tabular}{lllll}
  \hline
  type/length & 100 & 1000 & 10000 & 100000 \\ 
  \hline
 sin 0.1 & 45 (14) / 87 (1) & 13 (40) / 95 (1) & 28 (2) / 28 (1) & 24 (2) / 4 (3) \\ 
   sin 1 & 23 (8) / 18 (1) & 39 (21) / 29 (3) & 13 (58) / 10 (6) & 10 (90) / 38 (6) \\ 
  sin10 & 29 (7) / 25 (1) & 10 (90) / 9 (34) & 4 (96) / 7 (68) & 3 (97) / 6 (75) \\ 
  log 0.1 & 22 (7) / 5 (1) & 1 (1) / 4 (5) & 21 (7) / 6 (4) & 21 (1) / 58 (2) \\ 
   log 1 & 31 (1) / 16 (1) & 12 (8) / 21 (2) & 14 (17) / 27 (5) & 48 (46) / 28 (1) \\ 
  log10 & 42 (58) / 27 (10) & 65 (35) / 8 (11) & 4 (96) / 5 (30) & 15 (85) / 9 (63) \\ 
  lor 0.1 & 8 (2) / 3 (1) & 9 (1) / 10 (1) & 6 (2) / 21 (3) & 7 (1) / 29 (1) \\ 
  lor 1 & 70 (30) / 40 (4) & 65 (7) / 29 (12) & 3 (97) / 10 (31) & 2 (96) / 4 (45) \\ 
  lor 10 & 40 (57) / 30 (2) & 10 (15) / 4 (28) & 3 (97) / 3 (85) & 2 (98) / 3 (83) \\ 
   \hline
\end{tabular}
%\end{ruledtabular}
\end{table*}

%Namely, the data length, signal to noise ratio, the efficiency of computational algorithms.

\section{Discussion \label{s6}}

In  this study we have proposed a novel approach for performing a threshold free RQA and demonstrated its performance. The selection of the threshold can be avoided by averaging the RQA measure of interest which was calculated for a range of thresholds. 
We tested the ability of sorting data sets corrupted by white noise according their signal to noise ratios with the help of averaged DET measure.
The new approach 
performs more robust than standard single threshold approach.
 The explanation of the results  is that the  deterministic behaviour can be detected on more scales and provides a more robust RQA DET measure.

This property has been also achieved  when time series embedding is applied, as for the Lorenz system, where the embedding parameters have been set to 3 for the time lag as for the embedding dimension.

For the purpose of identifying deterministic components in noisy signals, the proposed approach might be the practical choice. The found RR value, out of this analysis, up to which the averaging should be performed  is $RR_* \approx 40 \%$, as in the vicinity of this value the maxima of the sorting rates were achieved for all the systems (Fig.~\ref{f5}). It also corresponds to previous findings that the discrimination of deterministic signals from noise works well for a quite large range of thresholds $\varepsilon$\cite{2008EPJST.164...45S}. Averaging to larger $RR_*$ might also work, but could reduce the robustness of the RQA measures.

However, we are aware of the fact that the complexity of the analysed artificial data is limited, and in the future further features could be introduced in order to explore the boundaries of this approach, namely gaps, other types of noise, different lengths of time series and sampling. The latter factors would help to better mimic the data obtained from unknown systems as they are the typical challenges in data analysis. Moreover, the suggested averaging approach was developed for the research question on discrimination a signal component from noisy signals, in particular, to order them with respect to the SNR. Whether it works also for other purpose should be studied in more detail in the future.

%Kritika approachu s jednym eps..

%Dolezitost tohoto vyskumu pre rozne typy data... 
%In astronomy where especiallt the data from the satelites are extremely expensive, one has the duty to squieze the most information from them. 

%Another kind of expensive data are the gravitational wave events, which dimensionality is with this new approach using GPU affordable even for..

\section*{Acknowledgments}
 RP acknowledges the institutional support of the Silesian University in Opava and the grant SGS/26/2022 and OP VVV $CZ.02.2.69/0.0/0.0/18\_053/0017871.$

\bibliographystyle{unsrtnat}
\bibliography{aRQA}

\begin{thebibliography}{26}
\providecommand{\natexlab}[1]{#1}
\providecommand{\url}[1]{\texttt{#1}}
\expandafter\ifx\csname urlstyle\endcsname\relax
  \providecommand{\doi}[1]{doi: #1}\else
  \providecommand{\doi}{doi: \begingroup \urlstyle{rm}\Url}\fi

\bibitem[{Bradley} and {Kantz}(2015)]{2015Chaos..25i7610B}
Elizabeth {Bradley} and Holger {Kantz}.
\newblock {Nonlinear time-series analysis revisited}.
\newblock \emph{Chaos}, 25\penalty0 (9):\penalty0 097610, September 2015.
\newblock \doi{10.1063/1.4917289}.

\bibitem[{Webber} and {Marwan}(2015)]{2015rqa..book.....W}
Jr. {Webber}, Charles~L. and Norbert {Marwan}.
\newblock \emph{{Recurrence Quantification Analysis}}.
\newblock Springer, 2015.

\bibitem[Marwan(2011)]{marwan2011}
N.~Marwan.
\newblock How to avoid potential pitfalls in recurrence plot based data
  analysis.
\newblock \emph{International Journal of Bifurcation and Chaos}, 21\penalty0
  (4):\penalty0 1003--1017, 2011.
\newblock \doi{10.1142/S0218127411029008}.

\bibitem[{Thiel} et~al.(2002){Thiel}, {Romano}, {Kurths}, {Meucci}, {Allaria},
  and {Arecchi}]{2002PhyD..171..138T}
Marco {Thiel}, M.~Carmen {Romano}, J{\"u}rgen {Kurths}, Riccardo {Meucci},
  Enrico {Allaria}, and F.~Tito {Arecchi}.
\newblock {Influence of observational noise on the recurrence quantification
  analysis}.
\newblock \emph{Physica D Nonlinear Phenomena}, 171\penalty0 (3):\penalty0
  138--152, October 2002.
\newblock \doi{10.1016/S0167-2789(02)00586-9}.

\bibitem[{Medrano} et~al.(2021){Medrano}, {Kheddar}, {Lesne}, and
  {Ramdani}]{2021Chaos..31h3131M}
Johan {Medrano}, Abderrahmane {Kheddar}, Annick {Lesne}, and Sofiane {Ramdani}.
\newblock {Radius selection using kernel density estimation for the computation
  of nonlinear measures}.
\newblock \emph{Chaos}, 31\penalty0 (8):\penalty0 083131, August 2021.
\newblock \doi{10.1063/5.0055797}.

\bibitem[{Andreadis} et~al.(2020){Andreadis}, {Fragkou}, and
  {Karakasidis}]{2020Chaos..30a3124A}
Ioannis {Andreadis}, Athanasios~D. {Fragkou}, and Theodoros~E. {Karakasidis}.
\newblock {On a topological criterion to select a recurrence threshold}.
\newblock \emph{Chaos}, 30\penalty0 (1):\penalty0 013124, January 2020.
\newblock \doi{10.1063/1.5116766}.

\bibitem[{Schinkel} et~al.(2008){Schinkel}, {Dimigen}, and
  {Marwan}]{2008EPJST.164...45S}
S.~{Schinkel}, O.~{Dimigen}, and N.~{Marwan}.
\newblock {Selection of recurrence threshold for signal detection}.
\newblock \emph{European Physical Journal Special Topics}, 164\penalty0
  (1):\penalty0 45--53, October 2008.
\newblock \doi{10.1140/epjst/e2008-00833-5}.

\bibitem[{Cencini} et~al.(2000){Cencini}, {Falcioni}, {Olbrich}, {Kantz}, and
  {Vulpiani}]{2000PhRvE..62..427C}
M.~{Cencini}, M.~{Falcioni}, E.~{Olbrich}, H.~{Kantz}, and A.~{Vulpiani}.
\newblock {Chaos or noise: Difficulties of a distinction}.
\newblock \emph{Physical Review E}, 62\penalty0 (1):\penalty0 427--437, July
  2000.
\newblock \doi{10.1103/PhysRevE.62.427}.

\bibitem[{Urry}(1996)]{1996ASPC..110..391U}
C.~M. {Urry}.
\newblock {An Overview of Blazar Variability}.
\newblock In H.~Richard {Miller}, James~R. {Webb}, and John~C. {Noble},
  editors, \emph{Blazar Continuum Variability}, volume 110 of
  \emph{Astronomical Society of the Pacific Conference Series}, page 391,
  January 1996.

\bibitem[Mohorian et~al.(2021)Mohorian, Bhatta, Adhikari, Dhital, Pánis,
  Dinesh, Chaudhary, Bachchan, and Stuchlík]{Mohorian_2021}
Maksym Mohorian, Gopal Bhatta, Tek~P Adhikari, Niraj Dhital, Radim Pánis,
  Adithiya Dinesh, Suvas~C Chaudhary, Rajesh~K Bachchan, and Zdeněk Stuchlík.
\newblock X-ray timing and spectral variability properties of blazars s5 0716 +
  714, {OJ} 287, mrk 501, and {RBS} 2070.
\newblock \emph{Monthly Notices of the Royal Astronomical Society},
  510\penalty0 (4):\penalty0 5280--5301, dec 2021.
\newblock \doi{10.1093/mnras/stab3738}.

\bibitem[Mayer et~al.(2010)Mayer, Kazantzidis, Escala, and
  Callegari]{Mayer2010}
L.~Mayer, S.~Kazantzidis, A.~Escala, and S.~Callegari.
\newblock {Direct formation of supermassive black holes via multi-scale gas
  inflows in galaxy mergers}.
\newblock \emph{Nature}, 466\penalty0 (7310):\penalty0 1082--1084, aug 2010.
\newblock ISSN 0028-0836.
\newblock \doi{10.1038/nature09294}.

\bibitem[{Francis} et~al.(1991){Francis}, {Hewett}, {Foltz}, {Chaffee},
  {Weymann}, and {Morris}]{1991ApJ...373..465F}
Paul~J. {Francis}, Paul~C. {Hewett}, Craig~B. {Foltz}, Frederic~H. {Chaffee},
  Ray~J. {Weymann}, and Simon~L. {Morris}.
\newblock {A High Signal-to-Noise Ratio Composite Quasar Spectrum}.
\newblock \emph{Astrophysical Journal}, 373:\penalty0 465, June 1991.
\newblock \doi{10.1086/170066}.

\bibitem[{Thorne}(1995)]{1995pnac.conf..160T}
K.~S. {Thorne}.
\newblock {Gravitational Waves}.
\newblock In E.~W. {Kolb} and R.~D. {Peccei}, editors, \emph{Particle and
  Nuclear Astrophysics and Cosmology in the Next Millenium}, page 160, January
  1995.

\bibitem[{Marwan} et~al.(2002){Marwan}, {Wessel}, {Meyerfeldt}, {Schirdewan},
  and {Kurths}]{2002PhRvE..66b6702M}
Norbert {Marwan}, Niels {Wessel}, Udo {Meyerfeldt}, Alexander {Schirdewan}, and
  J{\"u}rgen {Kurths}.
\newblock {Recurrence-plot-based measures of complexity and their application
  to heart-rate-variability data}.
\newblock \emph{Physical Review E}, 66\penalty0 (2):\penalty0 026702, August
  2002.
\newblock \doi{10.1103/PhysRevE.66.026702}.

\bibitem[Kraemer et~al.(2021)Kraemer, Datseris, Kurths, Kiss,
  {Ocampo-Espindola}, and Marwan]{kraemer2021}
K.~H. Kraemer, G.~Datseris, J.~Kurths, I.~Z. Kiss, J.~L. {Ocampo-Espindola},
  and N.~Marwan.
\newblock A unified and automated approach to attractor reconstruction.
\newblock \emph{New Journal of Physics}, 23:\penalty0 033017, 2021.
\newblock \doi{10.1088/1367-2630/abe336}.

\bibitem[{Zbilut} and {Webber}(1992)]{1992PhLA..171..199Z}
Joseph~P. {Zbilut} and Charles~L. {Webber}.
\newblock {Embeddings and delays as derived from quantification of recurrence
  plots}.
\newblock \emph{Physics Letters A}, 171\penalty0 (3-4):\penalty0 199--203,
  December 1992.
\newblock \doi{10.1016/0375-9601(92)90426-M}.

\bibitem[{Eckmann} et~al.(1987){Eckmann}, {Oliffson Kamphorst}, and
  {Ruelle}]{1987EL......4..973E}
J.~P. {Eckmann}, S.~{Oliffson Kamphorst}, and D.~{Ruelle}.
\newblock {Recurrence plots of dynamical systems}.
\newblock \emph{EPL (Europhysics Letters)}, 4:\penalty0 973, November 1987.
\newblock \doi{10.1209/0295-5075/4/9/004}.

\bibitem[Kraemer et~al.(2018)Kraemer, Donner, Heitzig, and Marwan]{kraemer2018}
K.~H. Kraemer, R.~V. Donner, J.~Heitzig, and N.~Marwan.
\newblock {Recurrence threshold selection for obtaining robust recurrence
  characteristics in different embedding dimensions}.
\newblock \emph{Chaos}, 28\penalty0 (8):\penalty0 085720, 2018.
\newblock \doi{10.1063/1.5024914}.

\bibitem[{Iwanski} and {Bradley}(1998)]{1998Chaos...8..861I}
Joseph~S. {Iwanski} and Elizabeth {Bradley}.
\newblock {Recurrence plots of experimental data: To embed or not to embed?}
\newblock \emph{Chaos}, 8\penalty0 (4):\penalty0 861--871, December 1998.
\newblock \doi{10.1063/1.166372}.

\bibitem[{Thieler} et~al.(2016){Thieler}, {Fried}, and {Rathjens}]{robper}
Anita~M. {Thieler}, Roland {Fried}, and Jonathan {Rathjens}.
\newblock {{RobPer}: An {R} Package to Calculate Periodograms for Light Curves
  Based on Robust Regression}.
\newblock \emph{Journal of Statistical Software}, 69\penalty0 (9):\penalty0
  1--36, 2016.
\newblock \doi{10.18637/jss.v069.i09}.

\bibitem[{Takens}(1981)]{1981LNM...898..366T}
Floris {Takens}.
\newblock {Detecting strange attractors in turbulence}.
\newblock In \emph{Lecture Notes in Mathematics, Berlin Springer Verlag},
  volume 898, page 366. 1981.
\newblock \doi{10.1007/BFb0091924}.

\bibitem[{Cao}(1997)]{1997PhyD..110...43C}
Liangyue {Cao}.
\newblock {Practical method for determining the minimum embedding dimension of
  a scalar time series}.
\newblock \emph{Physica D Nonlinear Phenomena}, 110\penalty0 (1):\penalty0
  43--50, February 1997.
\newblock \doi{10.1016/S0167-2789(97)00118-8}.

\bibitem[{Jiao} et~al.(2017){Jiao}, {Venkat}, and
  {Weissman}]{2017arXiv170405199J}
Jiantao {Jiao}, Kartik {Venkat}, and Tsachy {Weissman}.
\newblock {Mutual Information, Relative Entropy and Estimation Error in
  Semi-martingale Channels}.
\newblock \emph{arXiv e-prints}, art. arXiv:1704.05199, April 2017.

\bibitem[Schultz et~al.(2015)Schultz, Spiegel, Marwan, and
  Albayrak]{SCHULTZ2015997}
David Schultz, Stephan Spiegel, Norbert Marwan, and Sahin Albayrak.
\newblock Approximation of diagonal line based measures in recurrence
  quantification analysis.
\newblock \emph{Physics Letters A}, 379, 02 2015.
\newblock \doi{10.1016/j.physleta.2015.01.033}.

\bibitem[Rawald et~al.(2017)Rawald, Sips, and Marwan]{Rawald2017PyRQA}
Tobias Rawald, Mike Sips, and Norbert Marwan.
\newblock Pyrqa - conducting recurrence quantification analysis on very long
  time series efficiently.
\newblock \emph{Comput. Geosci.}, 104:\penalty0 101--108, 2017.

\bibitem[{Zbilut} et~al.(2002){Zbilut}, {Zaldivar-Comenges}, and
  {Strozzi}]{2002PhLA..297..173Z}
Joseph~P. {Zbilut}, Jos{\'e}-Manuel {Zaldivar-Comenges}, and Fernanda
  {Strozzi}.
\newblock {Recurrence quantification based Liapunov exponents for monitoring
  divergence in experimental data}.
\newblock \emph{Physics Letters A}, 297\penalty0 (3-4):\penalty0 173--181, May
  2002.
\newblock \doi{10.1016/S0375-9601(02)00436-X}.

\end{thebibliography}

\end{document}